\documentclass[a4paper, 10pt, twocolumn]{article}

\usepackage{amsfonts,amssymb,amsmath,amsthm}
\usepackage{subfigure}
\usepackage{graphicx}
\usepackage[footnotesize]{caption}
\usepackage[ruled, vlined]{algorithm2e}
\usepackage{color}
\renewcommand{\thepage}{}

\usepackage[top=1.5cm, left=1.5cm, right=1.5cm, bottom=1.5cm]{geometry}

\title{Primal-dual splitting scheme with backtracking for handling with epigraphic constraint and sparse analysis regularization.\vspace{-0.4cm}}

\author{Laurence Denneulin$^{1,2}$, Nelly Pustelnik$^{2,3}$, Maud Langlois$^1$, Ignace Loris$^{4}$ and \'Eric Thi\'ebaut$^{1}$.\\
\footnotesize $^1$Univ Lyon, Univ Lyon1, ENS de Lyon, CNRS, Centre de Recherche Astrophysique de Lyon UMR5574, F-69230 Saint-Genis-Laval, France\\\footnotesize  $^2$Univ Lyon, ENS de Lyon, Univ Claude Bernard Lyon 1, CNRS, Laboratoire de Physique,F-69342 Lyon, France\\\footnotesize 
$^3$ISPGroup \& INMA/ICTEAM, UCLouvain, Belgium\\\footnotesize 
$^4$D\'epartement de Math\'ematique, Universit\'e libre de Bruxelles, Boulevard duTriomphe, 1050 Bruxelles, Belgium
\vspace{-0.8cm}}
\date{\empty} 

\renewenvironment{abstract}{\bf\small {\em\ Abstract---}}{}

\begin{document}
\newcommand*{\V}[1]{\boldsymbol{#1}} 
\newcommand*{\M}[1]{\mathbf{#1}} 
\newcommand*{\Tag}[1]{{\text{#1}}} 
\newcommand*{\Op}[1]{\mathrm{#1}} 

\newcommand*{\Id}{\M{Id}}

\def\x{{\mathbf x}}
\def\L{{\cal L}}
\newcommand{\RR}{\mathbb{R}}
\newcommand*{\Ss}{\text{S}}
\newcommand*{\Is}{\text{I}}
\newcommand*{\Qs}{\text{Q}}
\newcommand*{\Us}{\text{U}}
\newcommand*{\is}{\text{I}_j}
\newcommand*{\qs}{\text{Q}_j}
\newcommand*{\us}{\text{U}_j}
\newcommand*{\Zs}{\text{Z}}
\newcommand*{\Js}{\text{J}}
\newcommand*{\Os}{\text{O}}
\newcommand*{\Vs}{\text{V}}

\newcommand*{\pix}{{\text{pix}}} 
\newcommand*{\pol}{{\text{p}}}   
\newcommand*{\unpol}{{\text{u}}} 

\newcommand*{\Iu}{I^{\text{u}}}
\newcommand*{\Ip}{I^{\text{p}}}
\newcommand*{\iu}{I^{\text{u}}_j}
\newcommand*{\ip}{I^{\text{p}}_j}
\newcommand*{\thetapix}{\theta_j}
\newcommand*{\Iue}{\widehat{I}^{\text{u}}}
\newcommand*{\Ipe}{\widehat{I}^{\text{p}}}

\newcommand*{\N}{\mathbb{N}}
\newcommand*{\Z}{\mathbb{Z}}
\newcommand*{\R}{\mathbb{R}}
\newcommand*{\C}{\mathbb{C}}

\newcommand*{\irm}{\mathrm{i}}
\newcommand*{\erm}{\mathrm{e}}
  
\newcommand*{\sro}{\sigma_{\text{ro}}}

\newcommand*{\Argmin}[1]{\underset{#1}{\text{Argmin}}\;}
\newcommand*{\argmin}[1]{\underset{#1}{\text{argmin}}\;}

\newcommand*{\vnorm}{\text{v}^{\text{norm}}}
\newcommand*{\vdiff}{\text{v}^{\text{diff}}}
\newcommand*{\vreal}{\text{v}^{\text{real}}}
\newcommand*{\Vnorm}{\text{V}^{\text{norm}}}
\newcommand*{\Vdiff}{\text{V}^{\text{diff}}}
\newcommand*{\Vreal}{\text{V}^{\text{real}}}

\newcommand*{\Fig}{Fig.~}
\newcommand*{\ie}{\textit{i.e.} }
\newcommand*{\eg}{\textit{e.g.} }
\newcommand*{\hide}[1]{}

\newcommand*{\Prox}{\text{prox}}
\let\origcdot=\cdot
\renewcommand*{\cdot}{\mathord{\,\mathchar"2201\,}}
%
\renewcommand{\thepage}{}
\maketitle

%

\begin{abstract} The convergence of many proximal algorithms involving a gradient descent relies on its Lipschitz constant. To avoid computing it, backtracking rules can be used. While such a rule has already been designed for the forward-backward algorithm (FBwB), this scheme is not flexible enough when a non-differentiable penalization with a linear operator is added to a constraint. In this work, we propose a backtracking rule for the primal-dual scheme (PDwB), and evaluate its performance for the epigraphical constrained high dynamical reconstruction in high contrast polarimetric imaging, under TV penalization.      
\end{abstract}

\vspace{-0.3cm}
\section{Introduction}
\label{sec:introduction}\vspace{-0.2cm}

The resolution of inverse problems remains a challenging task in image processing, especially when dealing with a large amount of data, such as in astrophysics (e.g. $10^6$ to $10^9$ pixels). Important advances have been made for handling non-differentiable objective function, thanks to proximal algorithmic schemes but an important issue is the impact on the convergence behaviour of the Lipschitz constant of the gradient. Yet, the calculus of this constant can be time consuming or difficult. To get round this issue, a backtracking rule can be used. Such a rule  has been designed for forward-backard iterations in~\cite{FISTA} but for many inverse problems forward-backward iterations are not flexible enough to handle complex regularization terms and/or constraints. We then need to resort to primal-dual schemes \cite{condat_primaldual_2013} for which we propose to design a backtracking rule.

Equipped with a backtracking rule for both forward-backward and primal-dual schemes we propose to evaluate the reconstruction performances of Total Variation (TV) \cite{rudin_nonlinear_1992} with standard regularization procedure considered in astrophysics that is hyperbolic Total Variation (TV-h) \cite{charbonnier_edge_preserving_1997} regularization. 

To evaluate the performance, we focus on high contrast polarimetric imagery which benefits in considering jointly a TV-based penalization and an epigraphic constraint. Indeed, if epigraphical constraint has been considered in polarimetric radio-interferometry \cite{repetti_non-convex_2017}, in  high contrast polarimetric direct imaging, the state-of-the-art does not take it in account \cite{van_holstein_polarimetric_nodate}.

Section~\ref{sec:first-section} introduces the notations and the objective function we are interested in. Section~\ref{sec:backtracking} presents the proposed backtracking rule for primal-dual proximal schemes and convergence results. Section~\ref{sec:experiment} provides the direct model considered in high contrast polarimetric imagery, provides some recalls on TV and TV-h as well as experimental comparisons.
 
\vspace{-0.2cm}
\section{Problem formulation}
\label{sec:first-section}\vspace{-0.2cm}
We denote by ${\V{x}} = ({\V{x}}_1,\ldots, {\V{x}}_{L})\in (\mathrm{R}^{N})^L$ the $L$-component signal of interest, each of size $N$.  Our goal is to estimate:\vspace{-0.3cm} \begin{equation}\vspace{-0.3cm}
\widehat{\V{x}} \in \Argmin{\V{x} \in ({\R^N})^L} \lbrace h(\V{x})+ \sum_{\ell=1}^L g_\ell (\Op{D}_\ell \V{x}_\ell) + \iota_C(\V{x}) \rbrace. 
\label{eq:Crit}
\end{equation} where $h: ({\R^N})^L \rightarrow ]-\infty,+\infty]$ is a convex and differentiable function with a $\beta$-Lipschitz gradient (may denote the data-fidelity term), $\forall \ell =\{1, \dots, L\}$, $\Op{D}_\ell \in \RR^{K_\ell \times N}$ denotes a linear operator, and $g_\ell: \R^{K_\ell} \rightarrow ]-\infty,+\infty]$ is a proper, lower semi-continuous (l.s.c.), convex function (may stands for the regularization term, including TV, as well as TV-h in the differentiable case). See \cite{Pustelnik_N_2016, denneulin_gretsi_2019} for an exhaustive list of penalization choices having this form. Finally, $\iota_C : ({\R^N})^L \rightarrow\R$ is an epigraphical constraint, written as:\vspace{-0.3cm} \begin{equation}\vspace{-0.3cm}
\label{eq:constraint}
C=\{\left(\V{x}_1, \dots, \V{x}_L\right)\! \in\!  \left( \RR^N \right)^L\;\vert\; \phi( \V{x}_{2}, \dots, \V{x}_{L}) \leq \V{x}_{1}\}
\end{equation} where $\phi$ proper, l.s.c and convex (cf. e.g. \cite{chierchia_epigraphical_nodate}).

\vspace{-0.2cm}
\section{Backtracking proximal primal-dual}\vspace{-0.2cm}
\label{sec:backtracking}
When $g_\ell$ is differentiable, forward-backward scheme, possibly with backtracking as in \cite{FISTA}, can be considered to estimate $\widehat{\V{x}}$. When $g_\ell$ is non-differentiable, a well adapted scheme is the primal-dual algorithm~\cite{condat_primaldual_2013}, whose main interest is to exploit the differentiability of $h$ and relies on proximal steps for $g_\ell$ and $\iota_C$. Setting $g(\M{D} \V{x} )= \sum_{\ell=1}^L g_\ell(\Op{D}_\ell \V{x}_\ell)$, the iterations are summarized in Algorithm~\ref{algo:PD}. The sequence $(\V{x}^{[t]})_{t\in \N}$ is insured to converge to $\widehat{\V{x}}$, if the following condition on the parameters $\tau^{[t]}, \sigma^{[t]} \geq 0$ involving the Lipschitz constant $\beta>0$ holds: \vspace{-0.2cm}\begin{equation}\vspace{-0.3cm}
1/\tau^{[t]} - \sigma^{[t]} \Vert \M{D} \Vert^2 \geq \beta/2.
\label{eq:condicondat}
\end{equation}\vspace{-0.8cm}\begin{algorithm}
			Set $\V{x}^{[0]} \in (\R^{N})^L$ and   $\V{y}^{[0]} \in \R^{K_1} \times \ldots \times \R^{K_L}$.\\
				\For{$t=0,1,\ldots$}{				Set $\tau^{[t]}, \sigma^{[t]} \geq 0$ such that (\ref{eq:condicondat}) holds.\\
				$\V{x}^{[t+1]}=  \Prox_{\tau^{[t]} \iota_C} \left( \V{x}^{[t]} - \tau^{[t]} \left(\nabla h({\V{x}}^{[t]}) + \M{D}^* \V{y}^{[t]} \right) \right)$;\\
				$\V{y}^{[t+1]}= \Prox_{\sigma^{[t]} g^*} \left(\V{y}^{[t]} + \sigma^{[t]} \M{D} \left(2\V{x}^{[t+1]}-\V{x}^{[t]}\right)\right)$.
				}
\caption{Primal-Dual (PD) Condat-V\~u algorithm}
\label{algo:PD}
\end{algorithm}\vspace{-0.6cm}

In the case where $\beta$ is unknown, we need to resort to backtracking scheme, whose proposed iterations are described in Algorithm~\ref{algo:PDwB}. The main idea is to start from a small estimate $\mathring{\beta}^{[0]}>0$ of $\beta$. Then at each iterations, to test whether the candidate $\mathring{\V{x}}^{[i]}$ yields a reduction of the majorant quadratic approximation of (\ref{eq:Crit}) tangent to the current iterate $\V{x}^{[t]}$, according to $\mathring{\beta}^{[i]}$. If the condition holds, $\beta^{[t]}$ is updated with $\mathring{\beta}^{[i]}$, else $\mathring{\beta}^{[i]}$ is increased. With such a condition, $\exists t \in \N$ such that $\beta^{[t]} \geq \beta$. The sequence $(\V{x}^{[t]})_{t \in \N}$ generated by Algorithm~\ref{algo:PDwB} thus converges to $\widehat{\V{x}}$.


\begin{algorithm}[!h]
Set $\V{x}^{[0]} \in (\R^{N})^L$ and   $\V{y}^{[0]} \in \R^{K_1} \times \ldots \times \R^{K_L}$, $\beta^{[0]}\geq 0$ and $ \eta >1$:\\
\For{$t=0,1,\ldots$}{
	\For{$i=0,1,\ldots$}{$\mathring{\beta}^{[i]}=\eta^i \beta^{[t]}$ and $\mathring{\tau}^{[i]}, \mathring{\sigma}^{[i]} \geq 0$ such that (\ref{eq:condicondat}) holds.\\
		$\mathring{\V{x}}^{[i]}=~ \Prox_{\mathring{\tau}^{[t]} \iota_C} \left(  \V{x}^{[t]} - \mathring{\tau}^{[i]} \left(\nabla h(\V{x}^{[t]}) + \M{D}^* \V{y}^{[t]} \right) \right)$;\\
		$\mathring{\V{y}}^{[i]}= \Prox_{\mathring{\sigma}^{[i]} g^*} \left(\V{y}^{[t]} + \mathring{\sigma}^{[i]} \M{D} \left(2\mathring{\V{x}}^{[i]}-\V{x}^{[t]}\right)\right)$;\\
		\If{$h(\mathring{\V{x}}^{[i]})\geq ~ h(\V{x}^{[t]})+\langle \mathring{\V{x}}^{[i]} - \V{x}^{[t]}, \nabla h (\V{x}^{[t]})\rangle$ $\quad \quad + \frac{\mathring{\beta}^{[i]}}{2} \Vert \mathring{\V{x}}^{[i]} - \V{x}^{[t]} \Vert^2$}{
		$\beta^{[t+1]}=\mathring{\beta}^{[i]}$, $\V{x}^{[t+1]}= \mathring{\V{x}}^{[i]}$ and $\V{y}^{[t+1]}= \mathring{\V{y}}^{[i]}$.\\\textbf{break}}
		}}
\caption{Primal-Dual with Backtracking (PDwB)}
\label{algo:PDwB}
\end{algorithm}

\vspace{-0.2cm}
\section{Experiments}
\label{sec:experiment}\vspace{-0.2cm}
\noindent \textbf{High contrast polarimetric imagery} --  We evaluate the performance of the  aglorithm \ref{algo:PDwB} to reconstruct circumstellar environments images using data from the Dual-Polarization Imaging (DPI) \cite{de_boer_polarimetric_nodate} modality of the SPHERE/IRDIS instrument \cite{beuzit_sphere:_2019,2014SPIE.9147E..1RL} installed at the Very Large Telescope (VLT) of the European Southern Observatory (ESO).

\begin{figure}[!b]\vspace{-0.3cm}
\includegraphics[width=0.5\textwidth]{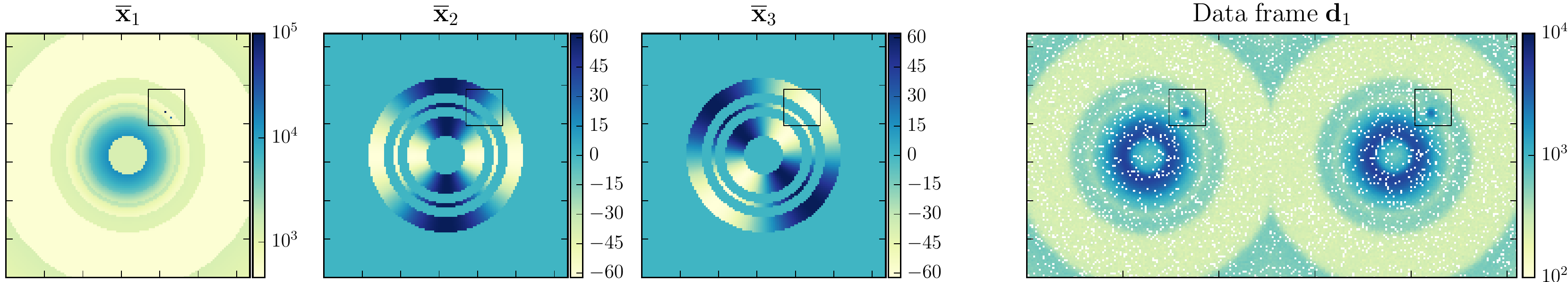} \vspace{-0.7cm}
\caption{True parameters and synthetics data.}
\label{fig:data}
\end{figure}

\noindent \textbf{Direct model} -- Observations consist in data cubes $\V{d} \in (\R^M)^K$ with $M=1024 \times 2048$ and $K$ a multiple of the four polarisation modulations in the instrument (\textit{e.g.} $K=64$ to $K>512$ depending on the object). The $L=3$ components to estimate (e.g. $\widehat{\V{x}}$) corresponds to three Stokes parameters $(\V{x}_1, \V{x}_2, \V{x}_3)=(\Is, \Qs, \Us)$, where $\Is$ is the total intensity while $\Qs$ and $\Us$ denote the linearly polarized intensity (resp. horizontal and vertical) \cite{van_holstein_polarimetric_nodate}. We created a synthetic object $\overline{\V{x}}$ (\textit{c.f.} Fig.~\ref{fig:data}) in order to be able to quantify the algorithmic performance. Synthetic data are created to be similar to real data (see Fig.~\ref{fig:data}). The dataset is composed of $K=64$ noise realizations following the direct model: \vspace{-0.4cm}\begin{equation}\vspace{-0.3cm}
(\forall k\in \{1, \dots K\}) \quad \V{d}_k=\mathcal{B} \left(  \sum_{\ell}  \left[ \begin{matrix} v_{k,\ell}^1 \M{A} \overline{\V{x}}_{\ell}\\ v_{k,\ell}^2 \M{A} \overline{\V{x}}_{\ell}  \end{matrix}\right] \right) 
\end{equation} \noindent where $\mathcal{B}(x)$ yields a realization of a Gaussian variable $\mathcal{N}(x, \mathrm{Diag}(x) +\sigma_\Tag{ro}^2\M{Id})$, to approximate Poisson noise plus read out noise of variance $\sigma_\Tag{ro}^2$, $\M{A} \in \RR^{N\times N}$ is the convolution with the PSF and the pairs $(v_{k, \ell}^1, v_{k, \ell}^2) \in \R^2$ represent polarization modulation at the acquisition $k$ on the $\ell$-th component. 

\noindent \textbf{Data-fidelity term $h$} --  It is the following Mahalanobis distance, such that, for every $\V{x} = (\V{x}_1,\ldots, \V{x}_{L})\in (\mathrm{R}^{N})^L$:
 \vspace{-0.2cm}\begin{equation}\vspace{-0.2cm}
 h\left({\V{x}}\right)= \sum_k \frac{1}{2} \left\Vert \V{d}_k - \sum_{\ell} \left[ \begin{matrix} v_{k,\ell}^1 \M{A} \V{x}_{\ell}\\ v_{k,\ell}^2 \M{A} \V{x}_{\ell}  \end{matrix}\right] \right\Vert_{\M{W}_\Tag{k}}^2,      
\end{equation} \noindent where  $\Vert \V{y} \Vert_{\M{W}_k}^2 = \V{y}^{\top}\M{W}_k \V{y}$ with $\M{W}_{k}= \textrm{Cov}(\V{d}_k)^{-1}$. This form of $h$ assumes that the $K$ data frames are mutually independent. 

\noindent \textbf{Epigraphical constraint $C$} -- The function $\phi$ in \eqref{eq:constraint} stems from the definition of the Stokes parameters and is given by:  \vspace{-0.3cm}\begin{equation}\vspace{-0.3cm}\forall n \in \{1, \dots, N\} \quad \phi(\V{x}_2, \dots, \V{x}_L)_n = \sqrt{\sum\nolimits_{\ell=2}^L \V{x}_{n,\ell}^2}.
\end{equation}It is important to avoid strong positive/negative oscillations that may result from the the deconvolution.

\noindent \textbf{Penalisation choice $g_\ell(\Op{D}_\ell \cdot)$: TV or TV-h} -- Unless brillant stars are in the field, circumstellar environments can be taken for piecewise constant objects. This motivate the use of edge-preserving penalization. We recall that TV is given $\forall \lambda_\ell \geq 0$ and $\forall \V{x} \in \mathcal{H}$, by: \vspace{-0.3cm}\begin{equation}\vspace{-0.2cm}
\Op{TV}_{\lambda_\ell}(\V{x}) = \lambda_\ell \Vert \nabla \V{x} \Vert_{\ell_1}. 
\label{eq:TV}
\end{equation} The formulation of TV-h is, $\forall \lambda_\ell \geq 0$, $\varepsilon > 0$ and $\forall \V{x} \in \mathcal{H}$: \vspace{-0.2cm}\begin{equation}\vspace{-0.2cm}
\Op{TV}_{\lambda_\ell,\varepsilon}^\Tag{h}(\V{x}) = \lambda_\ell \left(\sqrt{\Vert \nabla \V{x} \Vert_2^2 + \varepsilon^2} - \varepsilon \right). 
\label{eq:TVh}
\end{equation} 

\begin{figure}[!b]\vspace{-0.5cm}
\includegraphics[width=0.5\textwidth]{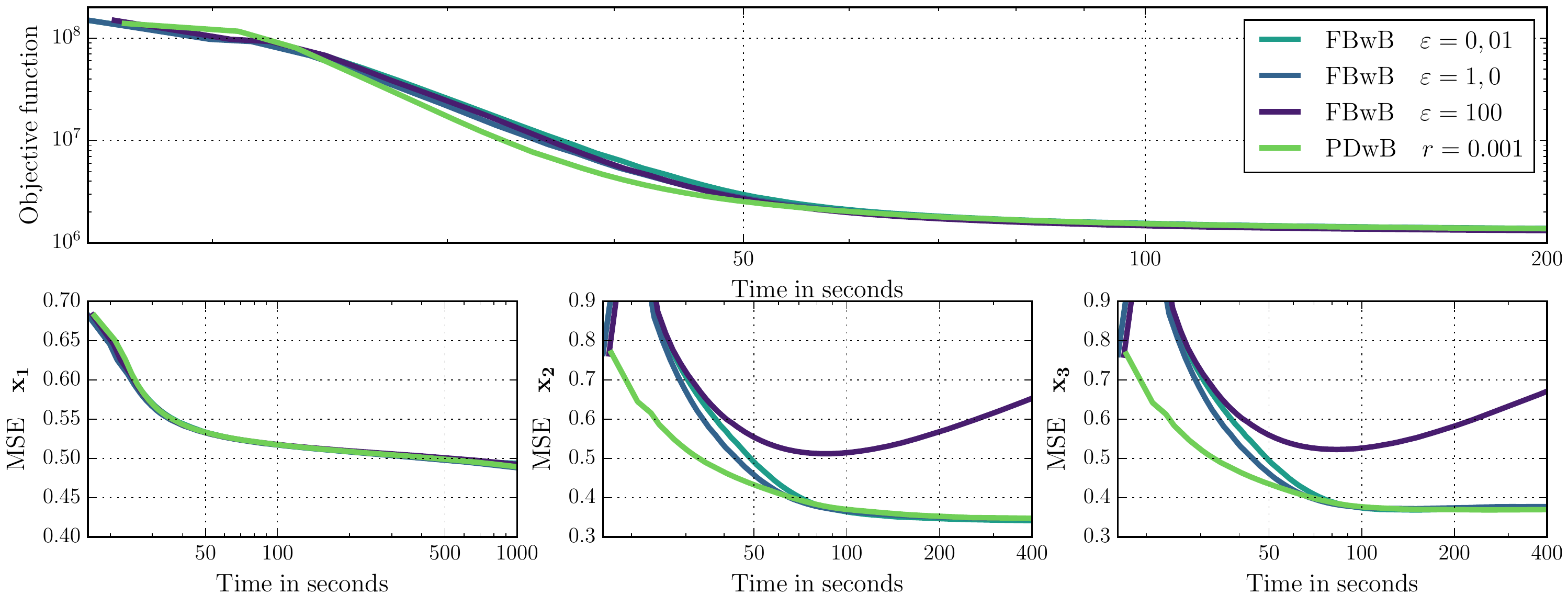} \vspace{-0.7cm}
\caption{Comparison of the convergence of the objective function and the Mean Square Error (MSE), as a function of the time in seconds.}
\label{fig:comp_conv} 
\end{figure}

\noindent \textbf{Performance evaluation} --   Figure~\ref{fig:comp_conv} shows the convergence of the objective function and of the normalized Mean Squared Error (MSE) of each $\ell$-th component, \textit{i.e.} $\Vert \widehat{\V{x}}_\ell- \overline{\V{x}}_\ell\Vert^2/\Vert \overline{\V{x}}_\ell \Vert^2$, as a function of the time. We compare the influence of the epigraphical constraint (\textit{i.e.} $\widehat{x} \in C$ and $\widetilde{x} \notin C$) on Fig.~\ref{fig:comp_conv}, on the high dynamical portion of $\V{x}$ highlighted on Fig.~\ref{fig:data}.

\noindent \textbf{Parameter selection} -- We performed the reconstruction with TV \cite{rudin_nonlinear_1992} and with TV-h \cite{charbonnier_edge_preserving_1997} for $\lambda_1 =0.1$ and $\lambda_2=\lambda_3=0,03$. For TV-h, we choose $\varepsilon$ in $\{10^{-2}, 1, 10^2\}$. We performed the TV-h reconstruction using the algorithm FBwB, with a descent step of $1.99/\beta^{[t]}$. We performed  the TV reconstruction using the algorithm \ref{algo:PDwB} with the parameters $\tau^{[t]}=(\beta^{[t]}/\gamma + r \Vert \M{D} \Vert^{2-s})^{-1}$ and $\sigma^{[t]}=r\Vert \M{D} \Vert^{-2}$, where $r >0$, $\gamma \in (0,2)$ and $s \in [0,2]$, inspired by the diagonal preconditioners proposed by Lorenz and Pock \cite[Lemma 10]{lorenz_inertial_2015} with $D=\beta \textrm{Id}$. We fixed $r=10^{-3}$, $\gamma=1.99$ and $s=2$ which seems to gives the fastest convergence. We started with $\beta^{[0]}=10^{-2}$ and set $\eta=1,1$.

 \begin{figure}[!b]\vspace{-0.3cm}
\includegraphics[width=0.5\textwidth]{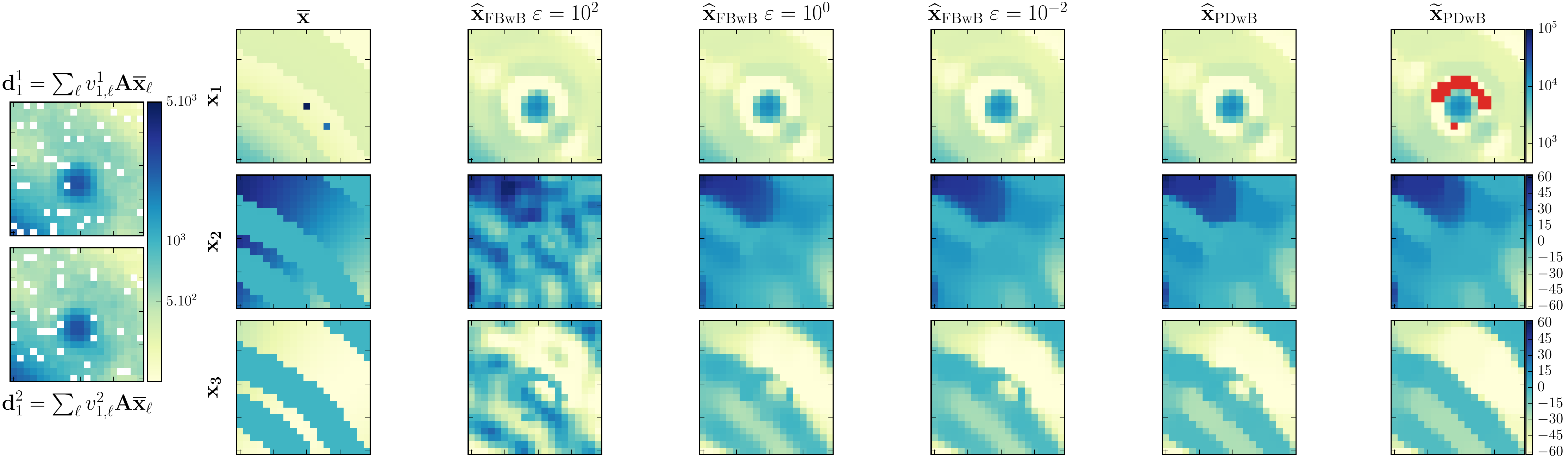} \vspace{-0.7cm}
\caption{Comparison of the reconstructed parameters $\widehat{\V{x}}$ for both methods and $\widetilde{\V{x}}$ for the PDwB algorithm. Pixels $n$ such that $\widetilde{\V{x}}_n \notin C$ are filled in red on $\widetilde{\V{x}}_1$.}
\label{fig:comp_visu}
\end{figure}

\noindent \textbf{Discussion} -- The epigraphical constraint reduces the oscillations around the two brillant dots (\textit{i.e.} stars) in $\widehat{\V{x}}_1$, yet it affects $\widehat{\V{x}}_3$. Without the epigraphical constraint, $\widetilde{\V{x}}_3$ is not affected by the deconvolution, yet the oscillations in $\widetilde{\V{x}}_1$ are amplified. In fact, the pixels of $\widetilde{\V{x}}_1$ filled with red on Figure~\ref{fig:comp_visu} are negatives. When no stars are in the field, the epigraphical constraint has no effects. It could thus be relaxed, in order to use differentiable methods with TV-h. In fact, TV and TV-h give similar results, unless $\varepsilon$ is large (\textit{i.e.} TV-h is mostly quadratic). However for the same time of convergence, TV still gives sharper edges than TV-h with $\varepsilon \rightarrow 0$. The choice of the method will then depend of the smoothness of the object. Finally, Figure~\ref{fig:comp_conv} validate numerically the PDwB algorithm. In fact, its convergences behaviour is similar to the convergence of FBwB, with TV-h for small values of $\varepsilon$.

\vspace{-0.2cm}
\section{Conclusion}\vspace{-0.2cm}
\label{sec:conclusion}

In this paper, we designed the PDwB algorithm, to handle both non-smooth TV and the epigraphical constraint. We applied PDwB to perform the reconstruction of simulated high dynamical images of circumstellar environments and compared the performances with FBwB using the TV-h. We observed that the backtracking is effective to achieve the convergence of primal-dual scheme when the Lipchitz constant is unknown, and that it could be applied for more complex reconstructions as texture decomposition. We observed that the epigraphical constraint is not always necessary, allowing the use of differential methods. 

\newpage

\end{document}